\documentclass[prl,twocolumn,letterpaper,10pt,twoside,tightenlines,nofootinbib,showpacs]{revtex4}
\usepackage{amsmath,amsfonts,amssymb,latexsym}
\usepackage{graphicx,xcolor}
\usepackage[sort&compress]{natbib}

\begin{document}

\newcommand{\eg}{{\it e.g.}}
\newcommand{\cf}{{\it cf.}}
\newcommand{\etal}{{\it et. al.}}
\newcommand{\ie}{{\it i.e.}}
\newcommand{\be}{\begin{equation}}
\newcommand{\ee}{\end{equation}}
\newcommand{\bea}{\begin{eqnarray}}
\newcommand{\eea}{\end{eqnarray}}
\newcommand{\bef}{\begin{figure}}
\newcommand{\eef}{\end{figure}}
\newcommand{\bce}{\begin{center}}
\newcommand{\ece}{\end{center}}
\newcommand{\red}[1]{\textcolor{red}{#1}}
\newcommand{\RR}[1]{\textcolor{purple}{\bf #1}}
\newcommand{\ZT}[1]{\textcolor{red}{#1}}
\newcommand{\dd}{\text{d}}
\newcommand{\ii}{\text{i}}
\newcommand{\lsim}{\lesssim}
\newcommand{\gsim}{\gtrsim}
\newcommand{\RAA}{R_{\rm AA}}
\newcommand{\QQb}{Q\bar{Q}}
\newcommand{\bbb}{b\bar{b}}

\title{Quarkonium Spectroscopy in the Quark-Gluon Plasma}

\author{Zhanduo~Tang$^{1}$, Biaogang~Wu$^1$, Andrew Hanlon$^{2}$, Swagato Mukherjee$^{3}$, Peter Petreczky$^{3}$, and Ralf~Rapp$^1$}
\affiliation{$^1$Cyclotron Institute and Department of Physics and
Astronomy, Texas A\&M University, College Station, TX 77843-3366, U.S.A. 
\\
$^2$Department of Physics, Kent State University, Kent, OH 44242, U.S.A.
\\
$^3$Physics Department, Brookhaven National Laboratory, Upton, NY 11973, U.S.A.}

\date{\today}

\begin{abstract}
{The properties of bound states are fundamental to hadronic spectroscopy and play a central role in the transition from hadronic matter to a quark-gluon plasma (QGP). In a strongly coupled QGP (sQGP), the interplay of temperature, binding energy and large collisional widths of the partons poses formidable challenges in evaluating the in-medium properties of hadronic states and their eventual melting. In particular, the existence of heavy quarkonia in the QGP is a long-standing problem that is hard to solve by considering their spectral properties on the real-energy axis. We address this problem by analyzing in-medium thermodynamic quarkonium $T$-matrices in the complex energy plane. We first validate this method in vacuum, where the $T$-matrix poles of observed states are readily identified.
When deploying this approach to recent self-consistently calculated $T$-matrices in the QGP, we find that poles in the complex energy plane can persist to surprisingly large temperatures, depending on the strength of the in-medium interactions. While the masses and widths of the pole positions are precisely defined, the notion of a binding energy is not due to the absence of thresholds caused by the (large) widths of the underlying anti-/quark spectral functions. Our method thus provides a new and definitive quantum-mechanical criterion to determine the melting temperature of hadronic states in the sQGP while increasing the accuracy in the theoretical determination of transport parameters.
 
}
\end{abstract}


\maketitle

{\it Introduction.---}
The search and characterization of phase transitions in Quantum Chromodynamics (QCD) matter is at the forefront of contemporary research. 
The changes in the properties of hadrons as the strongly interacting medium transits from confined and chirally broken hadronic matter into a deconfined and chirally restored quark-gluon plasma (QGP) play a pivotal role in this endeavor. While ultrarelativistic heavy-ion collisions (URHICs) have provided ample evidence for the formation of a QGP~\cite{Shuryak:2024zrh,Busza:2018rrf}, a microscopic description of its properties based on the underlying QCD forces remains a key challenge. Heavy quarkonia have been realized early on as a prime probe of the deconfinement transition~\cite{Matsui:1986dk}, due to their sensitivity to the ``confining force" in the heavy-quark (HQ) potential that is by now well established in the vacuum~\cite{Eichten:1979ms,Bali:2000gf}. 
The original idea of a suppression of quarkonia due to Debye screening has developed into a more complex picture which requires kinetic transport approaches based on inelastic reaction rates~\cite{Andronic:2024oxz}. The interplay of the reaction rates, representing the quarkonium decay widths, with the (in-medium) binding energies of the various quarkonium states is expected to govern their melting in the medium. At the same time, regeneration reaction will commence
once the cooling medium in the expanding fireball supports the reformation of the bound states. The notion of a ``melting temperature", which is of both practical and principal importance, plays a key role in this context.
Various prescriptions to define it have been proposed in the literature, \eg, the vanishing of the in-medium binding energy or the latter becoming comparable to the dissociation width, but a precise criterion has not been identified to date, cf.~Refs.~\cite{Rapp:2008tf,Braun-Munzinger:2009dzl,Kluberg:2009wc,Mocsy:2013syh,Liu:2015izf} for reviews.
All previous studies of quarkonia in the QGP are, in fact, restricted to either solving the Schr\"odinger equation or the analysis of spectral functions on the real-energy axis. Even though the latter treat binding energies and dissociation rates on the same footing, one is limited to the analysis of peak structures which are well defined only as long as their width remains reasonably small. However, as temperatures rise, the peaks become broader as the bound states lose binding energy and gradually merge into a continuum which makes a conclusive determination of a ``melting temperature" impossible.

In this paper, we propose a resolution to this issue by performing the first complex-pole analysis of in-medium $\QQb$ scattering amplitudes in the QGP. This method will enable to unambiguously determine whether a quarkonium exists in the QGP or has melted, by the presence or absence, respectively, of a pertinent pole in the complex energy plane of the in-medium $T$-matrix. This technique is well known in the analysis of bound-state structures in the vacuum~\cite{ParticleDataGroup:2024cfk,Briceno:2017max,Guo:2017jvc}, including multi-channel problems where the presence of thresholds leads to finite decay widths and thus shifts the location of bound-state poles into the complex plane. An in-medium $T$-matrix analysis has been conducted in Ref.~\cite{Albaladejo:2021cxj} in hadronic matter for the $D\bar{D}^*$ loop of a dynamically generated $X(3872)$ particle, using a constant effective (complex) mass for the $D$-meson selfenergies. Our method will account for the full off-shell properties of the single-particle propagators, which is particularly important in a strongly coupled quantum system with large collision widths. We will illustrate this method by carrying out quantitative analyses of bottomonium states using state-of-the-art $T$-matrix amplitudes which have been thoroughly constrained by recent lattice-QCD (lQCD) data~\cite{Tang:2023tkm} and are compatible with a strongly coupled QGP (sQGP) where the HQ potential is large and the collision rates (widths) of the heavy quarks are high. Alternatively, we will also analyze  a scenario with a substantially screened potential leading to a relatively weakly coupled QGP with small collision rates which is expected to cause a faster melting of bound states.
{\it Quantum Many-Body Approach.--}
Let us first recall the basic ingredients and features of the thermodynamic 
$T$-matrix formalism. It is based on a coupled set of 1- and 2-body correlation functions (propagators and scattering amplitudes) for heavy quarks ($Q$) and antiquarks ($\bar{Q}$) and their interactions in the QGP, schematically given by~\cite{Cabrera:2006wh,Riek:2010fk,Liu:2017qah}
\begin{eqnarray}
T_{Q\bar Q}(E,p,p') &=& V_{Q\bar Q}(p,p') + \int d^3k V_{Q\bar Q}(p,k) \nonumber \\
&&\times G_{Q\bar Q}(E,k)  T_{Q\bar Q}(E,k,p'),  \label{eq_Tm} 
\label{tmat}
\\
 G_{Q\bar Q}(E,k)&=&\int dk_0 G_{Q}(E-k_0,k)G_{\bar Q}(k_0,k),  \label{GQQ}
\\
G_{Q,i}(k_0,k) &=&  1/[k_0-\varepsilon_{Q,i}(k) -\Sigma_{Q,i}(k_0,k)] , 
\label{GQ}
\\
\Sigma_Q(k_0,k) &= & \int d^4p  T_{Qi}(k_0+p_0,k,p) 
G_{i}(p_0,p) n_i  \quad
\label{sigma}
\end{eqnarray}
where $T_{Q\bar Q}$ is the quarkonium $T$-matrix, $G_{Q,i}$ are the single-parton propagators for heavy quarks ($Q$) or thermal light partons ($i,j=q,\bar q,g$), and $G_{Q\bar{Q}}$ is the 2-body propagator from a convolution of HQ propagators. The on-shell energies, $\varepsilon_{Q,i}=\sqrt{M_{Q,i}^{2}+k^{2}}$, include in-medium particle masses $M_{Q,i}$
and energy-momentum dependent selfenergies, $\Sigma_{Q,i}$, which are computed self-consistently from the pertinent in-medium heavy-light and light-light $T$-matrices, $T_{Qj}$ and $T_{ij}$, respectively. 
The latter are based on the same color-interaction potential, $V$, as for heavy quarks (modulo relativistic corrections described below).
The latter is an input quantity constrained by lattice QCD. In the color-singlet channel we employ the following ansatz for the static potential~\cite{Megias:2005ve}
\begin{equation}
\widetilde{V}(r,T) = -\frac{4}{3} \alpha_{s} \left[\frac{e^{-m_{d} r}}{r} + m_{d}\right] -\frac{\sigma}{m_s} \left[e^{-m_{s} r-\left(c_{b} m_{s} r\right)^{2}}-1\right] \ ,
\label{Vstatic}
\end{equation}
which in vacuum reduces to the well-known Cornell potential,
$\widetilde{V}(r) = -\frac{4}{3} \frac{\alpha_{s}}{r}  +\sigma r$,  where 
$\alpha_s$=0.27 and $\sigma$=0.225\,GeV$^2$ are chosen to reproduce the vacuum free energy from lQCD~\cite{HotQCD:2014kol}. The parameters $m_d$ and $m_s$ represent Debye screening masses for the short-range color-Coulomb and long-range string interactions, respectively, and the parameter $c_b$ simulates in-medium string breaking at large distances. Upon Fourier transforming to momentum space, relativistic corrections~\cite{Riek:2010fk}
are included which also feature a vector component in the confining potential to better fit the empirical spin-induced splittings of charmonia and bottomonia in vacuum~\cite{Tang:2023lcn}. 
We also account for what has been referred to as an imaginary part of the potential~\cite{Laine:2006ns}, which results from 
an interference of inelastic parton scattering off the $Q$ and $\bar{Q}$ within
the quarkonium state. This reduces its dissociation width with decreasing radius until it vanishes for $r$$\to$0 (colorless limit). Diagrammatically, it corresponds to 3-body effects which, for simplicity, have been parameterized by a perturbatively inspired interference function~\cite{Laine:2006ns,Liu:2017qah,Tang:2023tkm}.

In the following we consider two scenarios to highlight the features of the complex-pole analysis and its sensitivity to the underlying micro-physics of $Q$-$\bar{Q}$ correlations in the QGP. In the first one, we employ state-of-the-art $T$-matrices that have been quantitatively constrained by Wilson line correlators (WLCs) from lQCD and are selfconsistently embedded into an equation of state that also agrees with the lattice~\cite{Tang:2023tkm}. These constraints require a strong input potential with rather little Debye screening, producing large collisional widths for both light and heavy partons, in excess of 0.5~GeV at all temperatures considered. This, in turn, leads to transport parameters which are in the range required by heavy-ion phenomenology. In the second scenario, the input potential is markedly screened which substantially weakens the in-medium potential and thus represents a ``weakly coupled scenario" (WCS)~\cite{Liu:2017qah}; its selfconsistent implementation still agrees with lQCD data for HQ free energies and the EoS. While its transport properties are disfavored by phenomenology, it will serve to exhibit differences in the bound-state properties borne out of the pole analysis compared to the first scenario.

\begin{figure}[!t]
\begin{minipage}{1.0\linewidth}
\centering
\includegraphics[width=0.9\textwidth]{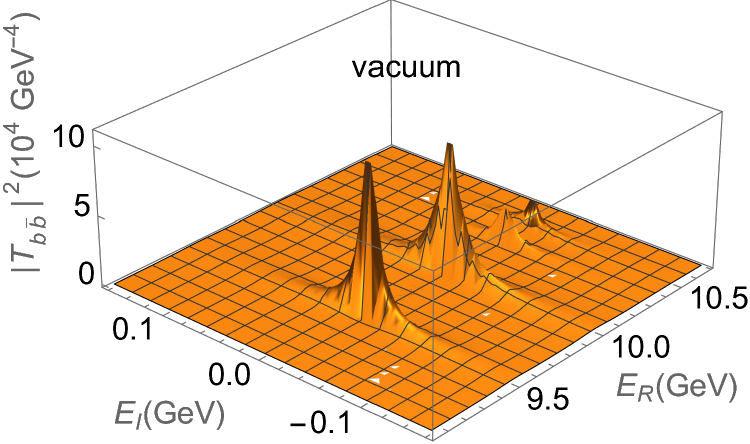}
\end{minipage}
\caption{The S-wave bottomonium $T$-matrix as functions of the real ($E_R$) and imaginary ($E_I$) energies at zero center-of-mass momentum in vacuum. For illustration purposes a small (artificial) $b$-quark width of 20\,MeV has been introduced in the 2-body propagator. All $\Upsilon$ states are below the vacuum $\bbb$ threshold of 10.513\,GeV.} 
\label{fig_Tm3D-vac}
\end{figure}

\begin{figure*}[htb]
\centering
\includegraphics[width=0.95\textwidth]{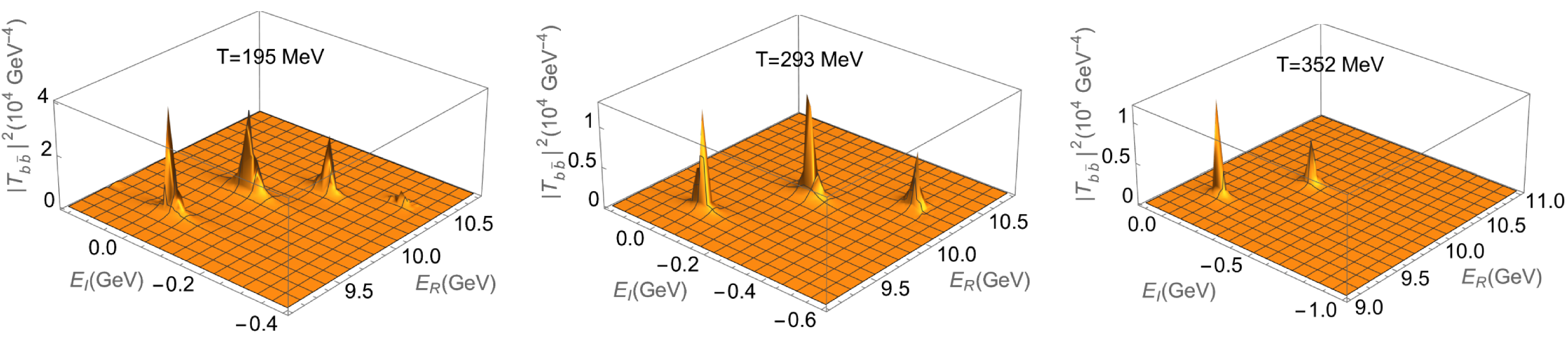}
\caption{The $S$-wave bottomonium $T$-matrices as functions of the real ($E_R$) and imaginary ($E_I$) energies at zero center-of-mass momentum at various temperatures for a strong potential constrained by WLCs, as following from our extension of the 2.~Riemann sheet to finite temperature.}
\label{fig_Tm3D_S_wlc}
\end{figure*}
%

{\it $T$-matrix at Complex Energies}.--
We are now proceed to the $T$-matrix analysis in the complex energy plane, thereby focusing on the case of bottomonia, as this system provides the largest mass scale ($m_b\simeq$\,5\,GeV).
We first note that the energy ($E$) dependence of the $T$-matrix, Eq.(\ref{tmat}), is generated entirely by the 2-body propagator. In the limit of small widths one can carry out the energy integration to obtain
\be
G_{b\bar{b}}(E,k)=\frac{1}{E\pm i\epsilon-2\varepsilon_{b}(k)},
\ee
where the $+i\epsilon$ case corresponds to the standard physical sheet (first Riemann sheet) with ${\rm Im} G_{b\bar{b}} < 0$, while the $-i\epsilon$ case is usually referred to as the second Riemann sheet and contains the information on bound states~\cite{Roca:2005nm,Gamermann:2006nm}. 
To access this information in the QGP, we define the complex 2-body selfenergy, 
$\Sigma_{b\bar{b}}(E,k)$, from the in-medium propagator in  Eq.~(\ref{GQQ}) via~\cite{Liu:2017qah}
\be
G_{b\bar{b}}(E,k)=\frac{1}{E-2\varepsilon_{b}(k) -\Sigma_{b\bar{b}}(E,k)} \ .
\label{Sigmabb}
\ee
That is, we first compute the 2-body propagator $G_{b\bar{b}}$ using Eq.~(\ref{GQQ}), and then obtain the 2-body selfenergy from Eq.~(\ref{Sigmabb}).
We then extend this expression into the complex energy plane by   
$E\to z=E_R - i E_I $ with $E_I>0$, \ie, 
\be
G_{b\bar{b}}(z,k)=\frac{1}{z-2\varepsilon_{b}(k) -\Sigma_{b\bar{b}}(E_R,k)} 
\ . 
\ee
For $E_I\to 0$ we recover the usual second Riemann sheet in the complex plane. 
By restricting the energy argument of $\Sigma_{b\bar{b}}$ to its real part, we ensure that the former reflects the physical properties of the $Q\bar Q$ states. In essence, the imaginary part of $z$ ``probes" the imaginary part of the selfenergy.
To make this more explicit, we can write the solution of the $T$-matrix equation in operator form as
\be
T(z) = \frac{V}{1- G_2(z) V}  \ , 
\ee
where the denominator is referred to as ``$T$-matrix determinant" or Jost function.
Its real part corresponds to a ``gap equation", which develops a zero if the potential is strong enough; large imaginary parts in the selfenergy of the 2-body propagator, $G_2$, tend to suppress possible solutions. On the other hand, for the imaginary part of the determinant to vanish, the imaginary part of $z$ has to compensate the value of $\mathrm{Im}\Sigma_{b\bar{b}} < 0$. 
A pole of $T(z)$  signals a quarkonium state thereby quantifying its mass and width, $\Gamma_Y$=$2E_I^{\rm pole}$. 
In the following, we utilize this method to examine the pole structure of bottomonia from selfconsistent in-medium $T$-matrix calculations as described above.

Before conducting the finite-temperature analysis, we benchmark our procedure by investigating the vacuum $T$-matrix, calculating its  absolute values in the complex energy plane, cf.~Fig.~\ref{fig_Tm3D-vac} for $S$-wave states. To better visualize the results, we adopt a small $b\bar b$ selfenergy of $\mathrm{Im}\Sigma_{b\bar b}$$=$$-20$ MeV in the propagator, shifting the bound-state positions to $E_I=-20$ MeV and generating  peak widths of $-2\mathrm{Im}\Sigma_{b\bar b}=40$ MeV.
We recover our previously identified $\Upsilon(1S)$, $\Upsilon(2S)$,
$\Upsilon(3S)$. In addition, we find a pole corresponding to the $\Upsilon(4S)$, which is located very close to the vacuum $b\bar{b}$ threshold and was not discernible from the real axis of the spectral functions in our earlier work~\cite{Tang:2023lcn}, \ie, even in vacuum the pole analysis is superior to using spectral functions.

\begin{figure*}[htb]
\centering
\includegraphics[width=0.95\textwidth]{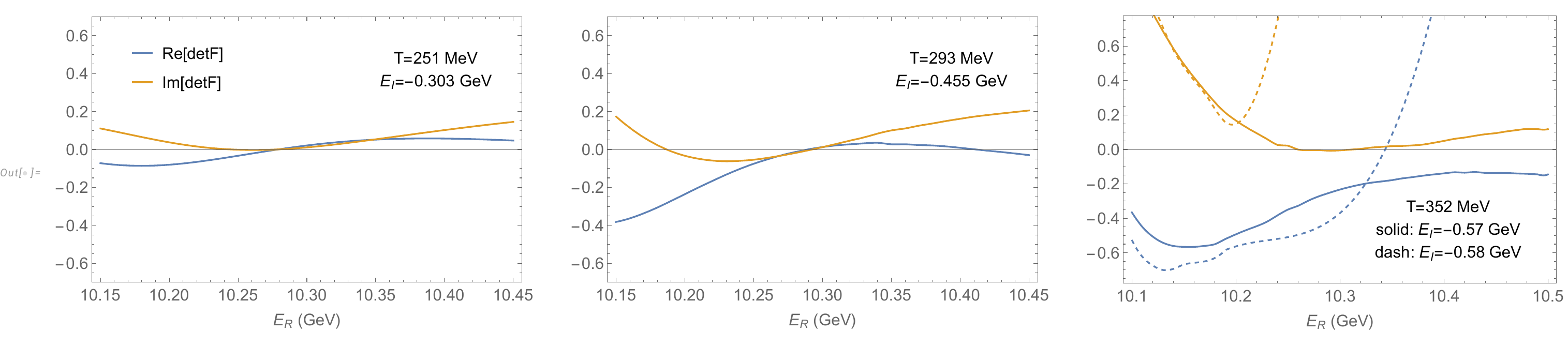}
\caption{Real and imaginary parts of Jost functions for the $\Upsilon(3S)$ vs.~$E_R$ at fixed $E_I$ based on Fig.~\ref{fig_Tm3D_S_wlc}; no more pole exists at $T$=352\,MeV,  with a melting near $T\simeq$\,330\,MeV.}
\label{fig_detF_3S}
\end{figure*}

{\it Bottomonium Spectroscopy in the QGP}.--
The absolute values of the in-medium $S$-wave $T$-matrices for bottomonia 
with the WLC-constrained potential (with weak screening even at relatively large temperatures)~\cite{Tang:2023tkm}  are displayed in the complex energy plane in Fig.~\ref{fig_Tm3D_S_wlc}. At the lowest temperature considered, $T$=195\,MeV, one recognizes the 3 poles corresponding to the $\Upsilon(1S)$, $\Upsilon(2S)$ and $\Upsilon(3S)$. In the QGP, the $\bbb$ threshold, 2$m_b(T)$, is no longer well defined due to large widths of the underlying $b$-quark spectral function. For a qualitative 
assessment of the states' ``binding energy" (which is no longer well defined either), we adopt an operational definition of a ``nominal" $\bbb$ threshold 
by using an effective mass of $b$-quarks as the average over their spectral function at each temperature.\footnote{We note that this deviates from the standard nomenclature in vacuum spectroscopy~\cite{Guo:2017jvc,Briceno:2017max} where {\em any} state above the lowest mass threshold is referred to as ``resonance".} With this caveat, we find that at $T$=195\,MeV, the $\Upsilon(1S)$ and $\Upsilon(2S)$ are still bound, with nominal binding energies (much) larger than their dissociation widths. However, the $\Upsilon(3S)$ width of $\sim$0.35\,GeV is already well above its nominal binding energy of below 0.1\,GeV. Even the $\Upsilon(4S)$ still persists, as a state above the nominal $\bbb$ threshold with a large width of $\sim$0.6\,GeV. 
This implies that there is no principal difference between a bound and a resonance state in the QGP, while a {\rm state} can still be unambiguously identified in the complex plane with a precisely defined mass and width. With increasing temperature, the $\Upsilon(4S)$ vanishes at $T$$\simeq$250\,MeV, and the $\Upsilon(3S)$ follows suit at $T$$\simeq$330\,MeV. The $\Upsilon(2S)$ survives up to $T$$\approx$700\,MeV, while the $\Upsilon(1S)$ persists to still higher temperatures. 

The vanishing of a pole can be illustrated more explicitly in a 1D projection, cf.~Fig.~\ref{fig_detF_3S} for the case of the $\Upsilon(3S)$. At the two lower temperatures, its pole is signified by a simultaneous zero in the real and imaginary part of the $T$-matrix determinant. Note that a second zero crossing in the real part does not correspond to a pole since the imaginary part does not vanish (and vice versa). The disappearance of the pole is realized when the zeros in the real part merge into a (single) maximum which drops below the zero line (the imaginary part still vanishes for a pertinent value of $E_I$), \ie, we are observing the vanishing of an isolated pole, not a pair of poles, due to the additional requirement on the imaginary part. The latter is thus at the origin of the single-pole mechanism. We have also verified that the residue of the pole diminishes as the temperature increases toward the melting point, using both the residue theorem in a closed contour around the pole and a matching of the $T$-matrix to a functional form of type $r_p/(z-z_{\rm pole})$, see, \eg, Ref.~\cite{Tiator:2010rp}.

\begin{figure}[!t]
\begin{minipage}{1.0\linewidth}
\centering
\includegraphics[width=0.99\textwidth]{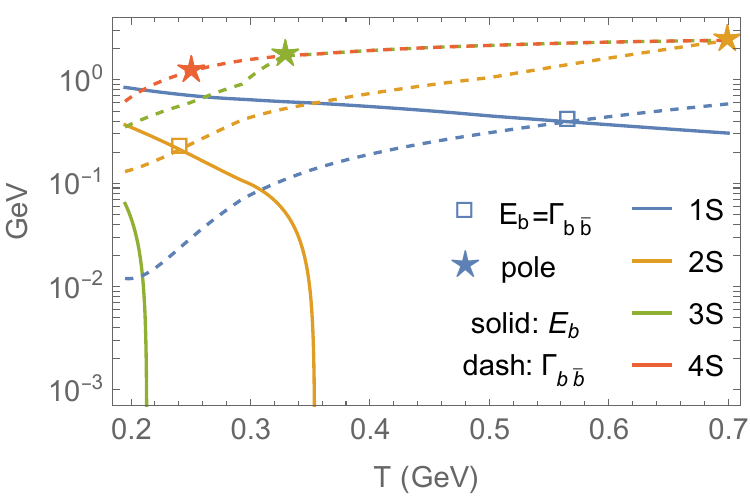}
\end{minipage}
\caption{Illustration of different melting criteria for $S$-wave bottomonia in the QGP. Solid line and dashed lines: nominal ``binding energies" (see text) and decay widths, respectively, extracted from the $T$-matrix poles with WLC constraints; stars indicate the vanishing of the pole and open boxes the temperature where the width equals the binding energy.} 
\label{fig_diss}
\end{figure}

The melting temperatures extracted from our pole analysis are put into context with previously used schematic criteria based on nominal binding energies and dissociation widths in Fig.~\ref{fig_diss}. Much higher melting temperatures are obtained from the pole analysis compared to previous estimates. 
In how far these features are related to recent lattice-QCD studies~\cite{Larsen:2019bwy}, where the use of extended-meson operators tailored to the spatial size of the vacuum states leads to rather well-defined peaks in pertinent spectral functions at relatively large temperatures, remains to be understood.

\begin{figure*}[!tb]
\centering
\includegraphics[width=0.95\textwidth]{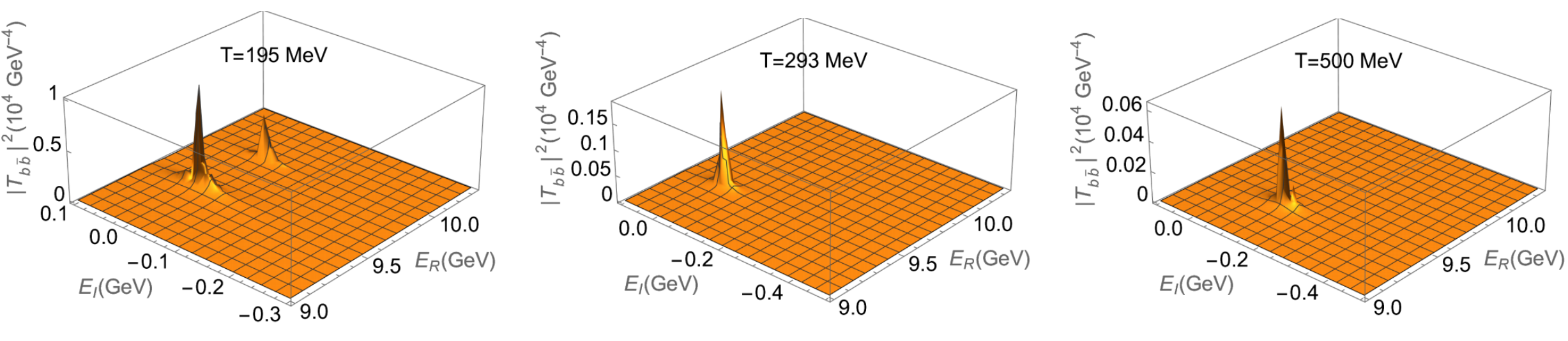}
\caption{S-wave bottomonium $T$-matrices as a function of real ($E_R$) and imaginary ($E_I$) energies at zero center-of-mass momentum at various temperatures for a weak potential constrained by HQ free energies~\cite{Liu:2017qah}.}
\label{fig_Tm3D_S_wcs}
\end{figure*}
Turning to the ``weakly coupled scenario" (constrained by the HQ free energy~\cite{Liu:2017qah}), we show  the results for $S$-wave bottomonium $T$-matrices displayed in Fig.~\ref{fig_Tm3D_S_wcs}. The much stronger screening of the underlying potential translates into a much reduced binding, while at the same time the collisional parton widths are smaller by up to a factor of $\sim$5 than those emerging from the WLC constraints. The $3S$ and $4S$ poles have already vanished at the lowest temperature, followed by the $2S$ at $T$=293\,MeV while the ground state survives to higher temperatures, albeit with reduced pole strength and a much reduced dissociation widths compared to the WLC constrained scenario. 
For this potential, states with vanishing nominal binding energies are not supported, suggesting that the latter are a characteristic of a strongly coupled system. However, the qualitative feature of a subsequent vanishing of single poles persists.

{\it Summary.--}
We have proposed a novel analysis to investigate ``bound-state" properties in medium by analyzing pertinent two-body $T$-matrices in the complex energy plane. By extending the energy variable in a suitably defined two-body propagator that retains the full-off-shell properties of the in-medium 1-body propagators, we are able to identify the in-medium poles of quark-antiquark states. This, in particular, allowed us to address the long-standing problem of the survival of quarkonia in the QGP, providing a definitive criterion for their existence and melting temperatures, while quantifying their masses and dissociation widths. We have found that the melting of individual states is realized  by the disappearance of a single pole in both the real and imaginary part of the $T$-matrix, accompanied by a decreasing residue prior to melting.

We have deployed this method using the self-consistent $T$-matrix  calculations for the bottomonium spectrum in the QGP. It turns out that the pertinent bound state poles persist much deeper into the QGP than expected from schematic energy uncertainty criteria. In a strongly coupled scenario with state-of-the-art $T$-matrices rooted in quantitative constraints from lattice QCD, we have found that bottomonia, despite large collisional widths of the $b$-quarks in a strongly coupled QGP, do not undergo a rapid melting, but continue to exist as broad states to much higher temperatures than the vanishing of their nominal binding energy would suggest, with dissociation widths reaching beyond 1\,GeV. On the other hand, in a weakly coupled scenario, the poles of all four $\Upsilon$ states vanish at markedly lower temperatures, signaling the sensitivity to the much stronger potential screening in this scenario. 
Elaborating the consequences of these findings for the phenomenology of heavy-ion collisions, in particular the quantitative sensitivity of observables to the persistence of hadronic states in the sQGP, will be an important objective for future transport studies. But the implications of our work go beyond the realm of quarkonia.
The question of degrees of freedom in the QGP equation of state and the transition back into a hadronic medium are at the core of the physics of a strongly-coupled quantum system which is driven by an interplay of large collisional widths and emerging composite states. Similar issues arise for hadronic matter, \eg, in recent discussions of whether light atomic nuclei can survive (and form) at temperatures which are much larger than their binding energies~\cite{Braun-Munzinger:2018hat}.
Our study has revealed that the interplay of the different scales can lead to rather unexpected results.
 

{\it Acknowledgments.--}
This work has been supported by the U.S. National Science Foundation under grant no. PHY-2209335, by the U.S. Department of Energy, Office of Science, Office of Nuclear Physics through contract No. DE-SC0012704 and the Topical Collaboration in Nuclear Theory on \textit{Heavy-Flavor Theory (HEFTY) for QCD Matter} under award no.~DE-SC0023547.

\bibliography{bibliography}

\end{document}